\documentclass[aip,jcp,reprint,noshowkeys,superscriptaddress]{revtex4-1}
\usepackage{graphicx,xcolor,physics,siunitx,xspace}
\usepackage{orcidlink}
\usepackage[version=4]{mhchem}
\usepackage[utf8]{inputenc}
\usepackage[T1]{fontenc}

\usepackage{svg}
\usepackage{algorithm2e}

\usepackage{hyperref}
\hypersetup{
    colorlinks,
    linkcolor={red!50!black},
    citecolor={red!70!black},
    urlcolor={red!80!black}
}

\DeclareSIUnit{\bohr}{\text{\ensuremath{a_{0}}}}
\DeclareSIUnit{\hartree}{\text{\ensuremath{E_{\textup{h}}}}}

\usepackage{listings}
\definecolor{codegreen}{rgb}{0.58,0.4,0.2}
\definecolor{codegray}{rgb}{0.5,0.5,0.5}
\definecolor{codepurple}{rgb}{0.25,0.35,0.55}
\definecolor{codeblue}{rgb}{0.30,0.60,0.8}
\definecolor{backcolour}{rgb}{0.98,0.98,0.98}
\definecolor{mygray}{rgb}{0.5,0.5,0.5}

\definecolor{sqred}{rgb}{0.85,0.1,0.1}
\definecolor{sqgreen}{rgb}{0.25,0.65,0.15}
\definecolor{sqorange}{rgb}{0.90,0.50,0.15}
\definecolor{sqblue}{rgb}{0.10,0.3,0.60}

\lstdefinestyle{mystyle}{
    backgroundcolor=\color{backcolour},
    commentstyle=\color{codegreen},
    keywordstyle=\color{codeblue},
    numberstyle=\tiny\color{codegray},
    stringstyle=\color{codepurple},
    basicstyle=\ttfamily\footnotesize,
    breakatwhitespace=false,
    breaklines=true,
    captionpos=b,
    keepspaces=true,
    numbers=left,
    numbersep=5pt,
    numberstyle=\ttfamily\tiny\color{mygray},
    showspaces=false,
    showstringspaces=false,
    showtabs=false,
    tabsize=2
  }

  \newcolumntype{d}{D{.}{.}{-1}}

\newcommand{\LCPQ}{Laboratoire de Chimie et Physique Quantiques (UMR 5626), Universit\'e de Toulouse, CNRS, UPS, France}
\newcommand{\Vienna}{%
  Institute for Theoretical Physics, TU Wien, %
  Wiedner Hauptstra{\ss}e 8--10/136, 1040 Vienna, Austria 
}

\begin{document}

\title{Stochastically accelerated perturbative triples correction in coupled cluster calculations}

\author{Yann \surname{Damour}\,\orcidlink{0000-0002-8868-9863}}
    \affiliation{\LCPQ}
\author{Alejandro \surname{Gallo}\,\orcidlink{0000-0002-3744-4161}} 
    \affiliation{\Vienna}
\author{Anthony \surname{Scemama}\,\orcidlink{0000-0003-4955-7136}}
    \email{scemama@irsamc.ups-tlse.fr}
    \affiliation{\LCPQ}

\begin{abstract}
We introduce a novel algorithm that leverages stochastic sampling
techniques to compute the perturbative triples correction in the
coupled-cluster (CC) framework.
By combining elements of randomness and determinism, our algorithm
achieves a favorable balance between accuracy and computational cost.
The main advantage of this algorithm is that it allows for
the calculation to be stopped at any time, providing an unbiased
estimate, with a statistical error that goes to zero as the exact
calculation is approached.
We provide evidence that our semi-stochastic algorithm achieves
substantial computational savings compared to traditional
deterministic methods.
Specifically, we demonstrate that a precision of 0.5 millihartree can
be attained with only 10\% of the computational effort required by the
full calculation.
This work opens up new avenues for efficient and accurate
computations, enabling investigations of complex molecular systems
that were previously computationally prohibitive.
\bigskip
\end{abstract}

\maketitle

\section{Introduction}
\label{sec:introduction}

Coupled cluster (CC) theory is an accurate quantum mechanical approach widely used in computational chemistry to describe the electronic structure of atoms and molecules.\cite{Cizek_1966,Cizek_1969,Paldus_1992}
In recent years, CC theories for both ground state and excited states
have received considerable attention in the context of material
science due to its good balance between accuracy and computational cost.\cite{APeriodicEquaGallo2021,GaussianBasedJames2017}
CC offers a systematic and rigorous framework for accurate predictions of molecular properties and reactions by accounting for electron correlation effects beyond the mean-field approximation.
The CC framework starts with a parameterized wave function, typically referred to as the CC wave function, which is expressed as an exponential series of particle-hole excitation operators acting on a reference state:
\begin{equation}
   \ket{\Psi_{\text{CC}}} = e^{\hat{T}} \ket{\Phi}
\end{equation}
where $\ket{\Phi}$ is the reference determinant, and $\hat{T}$ is the cluster operator representing single, double, triple, and higher excitations on top of the reference wave function.\cite{crawford_2000,bartlett_2007,shavitt_2009}

Coupled Cluster with Singles and Doubles (CCSD) includes single and double particle-hole excitations and represents the most commonly used variant of CC theory. CCSD is exact for two-electron systems and includes all terms
from third order perturbation theory and beyond.
Coupled Cluster with Singles, Doubles, and perturbative Triples (CCSD(T)) incorporates a perturbative correction to the CCSD energy to account for some higher-order correlation effects, and has been termed in the literature as the gold standard of quantum chemistry.\cite{raghavachari_1989}
CCSD(T) has demonstrated exceptional accuracy and reliability, making it one of the preferred choices for benchmark calculations and highly accurate predictions.
It has found successful applications in a diverse range of areas, including spectroscopy,\cite{villa_2011,watson_2016,vilarrubias_2020} reaction kinetics,\cite{dontgen_2015,castaneda_2012} and materials design,\cite{zhang_2019} and has played a pivotal role in advancing our understanding of complex chemical phenomena.

In the context of CC theory, the perturbative triples correction represents an important contribution to the accuracy of electronic structure calculations.\cite{stanton_1997}
However, the computational cost associated with the calculation of this correction can be prohibitively high, especially for large systems.
The inclusion of the perturbative triples in the CCSD(T) method leads to a computational scaling of $\order{N^7}$, where $N$ is proportional to the number of molecular orbitals.
This scaling can rapidly become impractical, posing significant challenges in terms of computational resources and time requirements.\cite{janowski_2008,deumens_2011,pitonak_2011,deprince_2013,anisimov_2014,peng_2019,shen_2019,gyevi_2020,wang_2020,datta_2021,gyevi_2021,jiang_2023}

To address this computational bottleneck, our goal is to develop a novel semi-stochastic algorithm that brings back the computational time to a level smaller or comparable to that of the CCSD method, which has a scaling of $\order{N^6}$, while ensuring well-controlled approximations.
Our algorithm strikes a balance between computational efficiency and
accuracy, making calculations for larger basis sets more feasible without compromising precision.
By incorporating stochastic sampling techniques, our approach provides an alternative avenue for approximating perturbative triples, relieving the computational burden inherent in traditional deterministic methods. This not only reduces the computational time to a more favorable level but also preserves the parallelism capabilities of CC calculations, ensuring efficient utilization of computational resources.

In the following sections, we will provide a brief introduction to the computation of perturbative triples in coupled cluster theory. We will explain the principles of our semi-stochastic algorithm, outlining its key features and advantages. Additionally, we will present implementation details, discussing the technical aspects and considerations involved in the algorithm's practical realization. To demonstrate the effectiveness and applicability of our approach, we finally present illustrative examples that showcase the algorithm's performance and compare it with the conventional algorithm.

\section{Theoretical Background}
\label{sec:theory}

The perturbative triples correction,
\begin{equation}
E_{(T)}  =  \sum_{ijk\,abc} E_{ijk}^{abc},
\end{equation}
is a sum of $N_{\text{o}}^3 \times N_{\text{v}}^3$ terms, where $N_{\text{o}}^3$ and  $N_{\text{v}}^3$ denote the number of occupied and virtual molecular orbitals, respectively.
For a closed-shell reference with canonical orbitals, each individual term is expressed as\cite{rendell_1991}
\begin{equation}
 E_{ijk}^{abc}  = \frac{1}{3} \frac{(4 W_{ijk}^{abc} +
              W_{ijk}^{bca} + W_{ijk}^{cab})
              (V_{ijk}^{abc} - V_{ijk}^{cba})}{\epsilon_i + \epsilon_j + \epsilon_k -
\epsilon_a - \epsilon_b - \epsilon_c},
\end{equation}
and depends on the canonical orbital energies $\epsilon$, and on the tensors $W$ and $V$:
\begin{align}
W_{ijk}^{abc} & = \Pi_{ijk}^{abc} \qty( \sum_d^{\text{virt}} \qty(bd|ai) t_{kj}^{cd} -
 \sum_l^{\text{occ}} \qty(ck|jl) t_{ab}^{il}) \\
V_{ijk}^{abc} & = W_{ijk}^{abc} + \qty(bj|ck) t_i^a + \qty(ai|ck) t_j^b + \qty(ai|bj) t_k^c
\end{align}
where $\Pi_{ijk}^{abc}$ is a sign-less permutation operator, $(t^a_i, t^{ab}_{ij})$ are the CCSD amplitudes,
\((pq|rs)\) are the two-electron coulomb integrals,
and the indices $i,j,k,l$ and $a,b,c,d$ refer to occupied and virtual orbitals, respectively.

The bottleneck of the perturbative triples correction is the computation of the $W$ tensor
which requires $\order{N_o^3 \times N_v^4}$ operations.
Fortunately, most of
the operations involved in the computation of $W$ can be recast into matrix
multiplications,\cite{form_w_abc} which are among the most efficient operations than can be
executed on modern CPUs and
accelerators.\cite{ma_2011,haidar_2015,dinapoli_2014,springer_2018}

In the algorithm proposed by Rendell\cite{rendell_1991}, for each given triplet $(a,b,c)$, the sub-tensors $W^{abc}$ and $V^{abc}$ are computed and immediately utilized to calculate their contribution to $E^{abc}$. Here, we propose a similar approach but introduce a semi-stochastic algorithm to randomly select the triplets $(a,b,c)$, circumventing the need to compute all contributions.

\section{Semi-Stochastic Algorithm}
\label{sec:algorithm}

\subsection{Stochastic formulation}

We propose an algorithm influenced by the semi-stochastic approach introduced in Ref.~\citenum{garniron_2017}, originally developed for computing the Epstein-Nesbet second-order perturbation correction to the energy. 

The perturbative triples correction is expressed as a sum of corrections, each indexed solely by virtual orbitals:
\begin{equation}
E_{(T)} = \sum_{abc} E^{abc} \text{, where }
E^{abc} = \sum_{ijk} E_{ijk}^{abc}.
\end{equation}
Monte Carlo sampling is employed by selecting samples $E^{abc}$.
The principal advantage of this formulation is that the number of triplet combinations $(a,b,c)$, given by $N_v^3$, is sufficiently small to allow for all contributions $E^{abc}$ to be stored in memory.
The first time a triplet $(a,b,c)$ is drawn, its corresponding value $E^{abc}$ is computed and then stored.
Subsequent drawings of the same triplet retrieve the value from memory. We refer to this technique as \emph{memoization}.
Thus, the computational expense of calculating the sample, which scales as $N_\text{o}^3 \times N_\text{v}$, is incurred only once, with all subsequent accesses being made at no cost.
Consequently, employing a sufficient number of Monte Carlo samples to ensure that each contribution is selected at least once results in a total computational cost that is only negligibly higher than that of an exact computation.

To reduce the fluctuations of the statistical estimator, we apply importance sampling: the samples are drawn using the probability
\begin{equation}
P^{abc} = \frac{1}{\mathcal{N}} \frac{1}{\max \left(\epsilon_{\min}, \epsilon_a + \epsilon_b + \epsilon_c \right)}
\end{equation}
where $\mathcal{N}$ normalizes the sum such that $\sum_{abc} P^{abc} = 1$, and $\epsilon_{\min}$ is an arbitrary minimal denominator to ensure that $P^{abc}$ does not diverge. In our calculations, we have set $\epsilon_{\min}$ to 0.2~a.u.
The perturbative contribution is then evaluated as an average over $M$ samples
\begin{equation}
E_{(T)} = \left\langle \frac{E^{abc}}{P^{abc}} \right \rangle_{P^{abc}} = 
          \lim_{M \to \infty} \sum_{abc} \frac{n^{abc}}{M} \frac{E^{abc}}{P^{abc}}.
\end{equation}
where $n^{abc}$ is the number of times the triplet $(a,b,c)$ was drawn with probability $P^{abc}$.

\begin{figure}[t]
\centering
\includegraphics[width=\columnwidth]{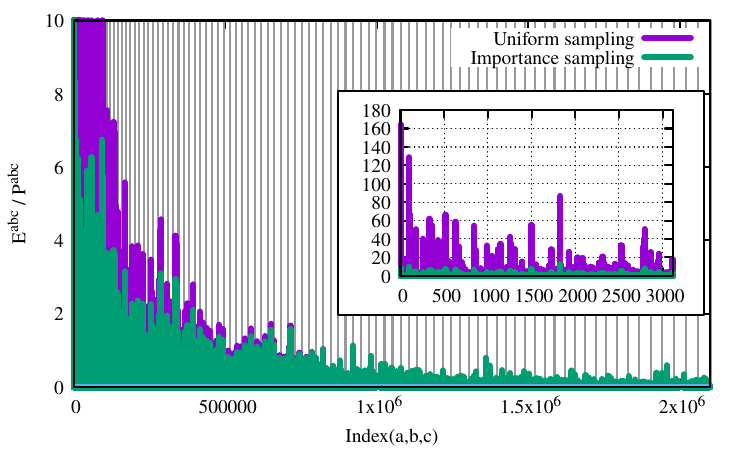}
\caption{%
Ratios $\frac{E^{abc}}{P^{abc}}$ obtained with the data of benzene/cc-pVTZ, using uniform or importance sampling.
Every bucket, delimited by vertical bars, contains a number of triplets such that the sum
\(\sum_{(a,b,c)}P^{abc}\) remains as uniform as possible. The zoomed window corresponds to the first bucket.
The fluctuations originating from the discrepancy of the values in the first buckets are considerably reduced by importance sampling.
\label{fig:buckets}
}
\end{figure}

This approach effectively reduces the statistical error bars by approximately a factor of two for the same computational expense due to two primary reasons: i) the estimator exhibits reduced fluctuations, ii) triplet combinations with low-energy orbitals are significantly more likely to be selected than others, enhancing the efficiency of memoization (see Fig.~\ref{fig:buckets}).

We employ the inverse transform sampling technique to select samples, where an array of pairs $\qty(P^{abc}, (a,b,c))$ is stored.
To further reduce the variance of the samples, this array is sorted in descending order based on $P^{abc}$ and subsequently partitioned into buckets as can be seen diagrammatically in Figure~\ref{fig:buckets}.
The partitioning into buckets is designed such that the sum $\sum_{(a,b,c) \in B} P^{abc}$ within each bucket $B$ is as uniform
as possible across all buckets.
As each bucket is equally probable, samples are defined as combinations of triplets, with one triplet drawn from each bucket.
Should the values of $E^{abc}$ be skewed, this advanced refinement significantly diminishes the variance.

The total perturbative contribution is computed as the aggregate of contributions from various buckets:
\begin{equation}
E_{(T)} = \sum_B E_B = \sum_B\sum_{(a,b,c) \in B} E^{abc}.
\end{equation}
Once every triplet within a bucket $B$ has been drawn at least once, the contribution $E_B$ can be determined.
At this juncture, there is no longer a necessity to evaluate \(E_B\) stochastically, and the buckets can be categorized into
deterministic ($\mathcal{D}$) and stochastic ($\mathcal{S}$) groups:
\begin{equation}
\label{eq:separation}
E_{(T)} = \sum_{B \in \mathcal{D}} E_B + \frac{1}{|\mathcal{S}|} \sum_{B \in \mathcal{S}}
\left \langle \frac{E^B_{abc}}{P^{abc}} \right \rangle_{P^{abc}}.
\end{equation}
Not all buckets are of equal size (see Figure~\ref{fig:buckets}); the number of triplets per bucket increases with the bucket's index. Consequently, the initial buckets transition into the deterministic set first, gradually reducing the stochastic contribution. When every triplet has been drawn, the exact value of $E_{(T)}$ is obtained, devoid of statistical error.
To accelerate the completion of the buckets, each Monte Carlo iteration triggers concurrently the computation of the first non-computed triplet. This ensures that after $N$ drawings, the
exact contribution from each bucket can be obtained.

The computational time required to generate a random number is negligible compared to the time needed to compute a contribution, $E^{abc}$.
Therefore, it is possible to obtain the exact contribution, characterized by zero statistical error, within a time frame equivalent to that required by a standard deterministic algorithm. This proposed algorithm offers the additional advantage of allowing the calculation to be terminated at any point prior to completion, with a statistical error.

\subsection{Implementation Details}
\label{sec:implementation}

\begin{algorithm}[H]
 \caption{\label{alg:stoch} Pseudo-code for the computation of the perturbative triples correction implemented in Quantum Package. $i_\text{min}$ denotes the first non-computed triplet, $w_{\text{accu}}$ contains the cumulative probability density, $\text{Search}(A, x)$ searches for $x$ in array $A$, $\text{First}(i)$ and $\text{Last}(i)$ return the first last indices belonging to bucket $i$.}
$i_{\text{min}} \leftarrow 1$ ;
$N^{abc}[1,\dots,N_{\text{triplets}}] \leftarrow [-1, -1, \dots]$ \;
$t_0 \leftarrow \text{WallClockTime}()$ \;
\For {$i_{\text{iter}}=1,\ldots,N_{\text{triplets}}$}
{
 \tcc{Deterministic computation}
 \While {$N^{abc}[i_{\text{min}}] > -1$ and $i_{\text{min}} \le N_{\text{triplets}}$}
 {
   $i_{\text{min}} \leftarrow i_{\text{min}}+1$ \;
 }

 \If{$i_{\text{min}} \le N_{\text{triplets}}$}
 {
   Send OpenMP task \{ {
     $E[i_{\text{min}}] \gets \text{Compute}(i_{\text{min}})$\;
     \} }\;
 }
 \tcc{Stochastic computation}

 $\eta \gets \text{RandomNumber}()$ \;
 \For {$i_\text{bucket} = 1,\ldots, N_{\text{buckets}}$}
 {
   \If{$i_{\text{min}} \le \text{Last}(i_\text{bucket})$}
   {
     $i_\eta \gets \text{Search}(w_{\text{accu}}, \frac{\eta + i_\text{bucket}-1}{N_{\text{buckets}}})+1$ \;
     \If{$N^{abc}[i_{\eta}] = -1$}
     {
       $N^{abc}[i_{\eta}] \gets 0$ \;
       Send OpenMP task \{ {
         $E[i_{\eta}] \gets \text{Compute}(i_{\eta})$\;
         \} }\;
     }
     $N^{abc}[i_{\eta}] \gets N^{abc}[i_\eta]+1$ \;
   }
 }
 \tcc{Compute the mean and error every second}
 $t_1 \gets \text{WallClockTime}()$ \;
 \If {$t_1 - t_0 > 1$ or $i_{\text{min}} \ge N_{\text{triplets}}$}
 {
   $i_\text{bucket} = 0$ \;
   \While {$i_{\text{bucket}} < N_\text{buckets}$ and
     $i_{\text{min}} > \text{Last}(i_\text{bucket}+1)$}
   {
     $i_\text{bucket} \gets i_\text{bucket} + 1$\;
   }
   $\mathcal{N} \gets \frac{ \sum_{i=\text{First}(i_\text{bucket}+1)}^{N_\text{triplets}} \max(N^{abc}[i],0)}
   { 1 - \sum_{i=1}^{\text{Last}(i_\text{bucket})} P[i] }$ \;
   $E_{d} \gets \sum_{i=1}^{\text{Last}(i_\text{bucket})} E^{abc}[i]$ \;
   $E_{s} \gets \frac{1}{\mathcal{N}}\, \sum_{i=\text{First}(i_\text{bucket}+1)}^{N_\text{triplets}}
   \max(N^{abc}[i],0)\, E^{abc}[i]/P[i]$ \;
   $E_{s^2} \gets \frac{1}{\mathcal{N}}\, \sum_{i=\text{First}(i_\text{bucket}+1)}^{N_\text{triplets}}
   \max(N^{abc}[i],0)\, \qty(E^{abc}[i]/P[i])^2$ \;
   $E \gets E_{d} + E_{s}$ \;
   $\Delta E \gets \sqrt{ \qty(E_{s^2} - {E_{s}}^2) / \qty(\mathcal{N}-1) }$ \;
   \If{$\Delta E < \epsilon$}
   {
     Exit outermost loop \;
   }
 }
}
\end{algorithm}

The algorithm presented in Algorithm~\ref{alg:stoch} was implemented in the \textsc{Quantum Package} software.
\cite{garniron_2019}
The stochastic algorithm is implemented using OpenMP tasks, where each task
consists in the computation of a single component $E^{abc}$.
The computation of the running average and statistical error is executed every second,
for printing or for exiting when the statistical error gets below a given threshold.

The number of samples $N^{abc}$ of each triplet $(a,b,c)$ is initialized to $-1$, to identify
the contributions that have not been already computed.
An outer \emph{for loop} runs over the maximum number of iterations, equal by construction to
the number of different triplets $N_{\text{triplets}}$.

Within a loop iteration, the index of the first non-computed triplet $(a,b,c)$ is identified, and the task associated with its computation is sent to the task queue.
As this triplet has never been drawn, $N^{abc}$ is set to zero.
Then, a triplet $(a,b,c)$ is drawn randomly.
If the $E^{abc}$ has not been computed (identified by $N^{abc}=-1$), the number of samples is set to zero and the task for the computation of this contribution is enqueued.
In any case, $N^{abc}$ is then incremented.

\subsection{Convergence of the statistical error in benzene}

In this section we illustrate the convergence of the statistical error of the perturbative triples correction as a function of the computational cost.
The benzene molecule serves as our reference system for conducting frozen-core CCSD(T) calculations employing the cc-pVTZ and cc-pVQZ basis sets.
Essentially, this involves the correlation of 30 (\(N_\text{o} = 15\)) electrons using either 258 (\(N_\text{v} = 243\)) or 503 (\(N_\text{v} = 488\)) molecular orbitals.
The calculations were performed on an AMD \textsc{Epyc} 7513 dual socket server (64 cores in total).

\begin{figure}[tb]
\includegraphics[width=\columnwidth]{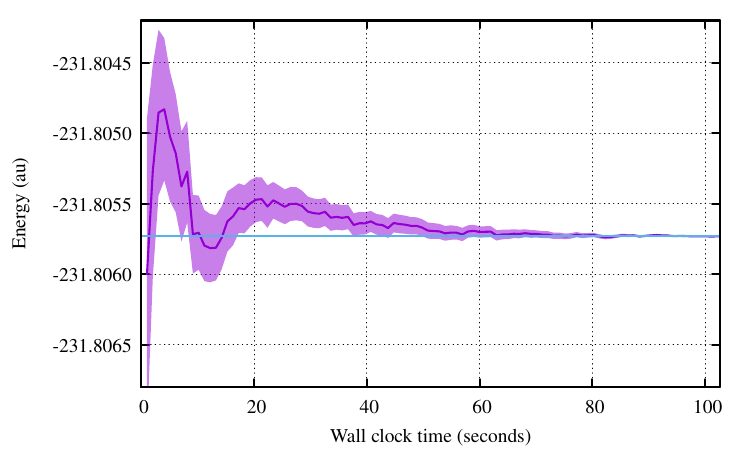}
\includegraphics[width=\columnwidth]{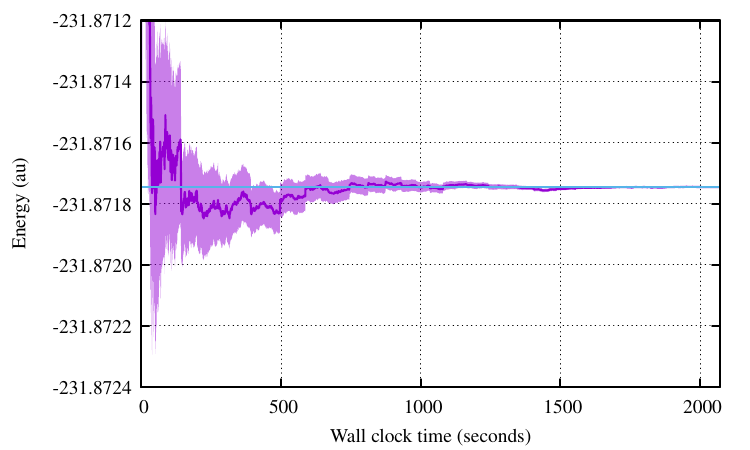}
    \caption{\label{fig:benzene} Energy convergence of benzene plotted against the program execution time, showing comparisons between the cc-pVTZ (upper curve) and cc-pVQZ (lower curve) basis sets. The blue lines indicate the exact CCSD(T) energies.}
\end{figure}

\begin{figure}[tb]
\includegraphics[width=\columnwidth]{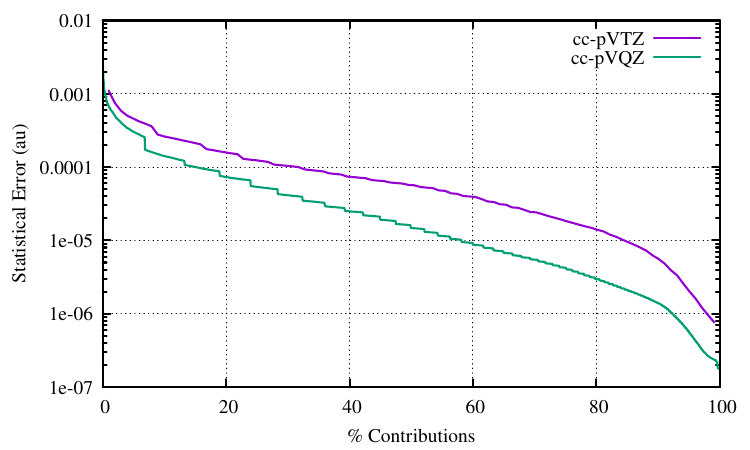}
    \caption{\label{fig:benzene_err} Convergence of the statistical error of the perturbative triples contribution in benzene as a function of the percentage of computed contributions, for both cc-pVTZ and cc-pVQZ basis sets.}
\end{figure}

Figure~\ref{fig:benzene} shows the convergence of the CCSD(T) energy as a function of the program execution time using the two basis sets.
Notably, the exact CCSD(T) energy always falls within $2\sigma$, affirming the reliability of the statistical error.
Figure~\ref{fig:benzene_err} displays the statistical error as a function of the percentage of computed contributions.
Noteworthy in the figure are the curve discontinuities, attributable to readjustments in the separation between the deterministic and stochastic components of the calculation (Eq.~\eqref{eq:separation}).
These updates lead to revised estimates and a diminution in statistical error.

Achieving chemical accuracy, defined as \SI{1.6}{\milli\hartree},\cite{pople_1999} necessitates less than 1\% of the total contributions in both basis sets.
Attaining a \SI{0.1}{\milli\hartree} precision level requires computation of 32\% and 15\% of the contributions for cc-pVTZ and cc-pVQZ, respectively.
The more rapid convergence observed with the larger basis set aligns with expectations, as expanding the basis set tends to increase the proportion of minor contributions while maintaining a relatively steady count of significant contributions.
This trend underscores the algorithm's enhanced suitability for systems with fewer electrons and extensive basis sets, as opposed to larger electron counts in smaller basis sets.

\subsection{Vibrational frequency of copper chloride}

Our methodology proves especially advantageous for scenarios requiring the
aggregation of numerous CCSD(T) energies, such as neural network training or
the exploration of potential energy surfaces.
In a recent article, Ceperley \textit{et al} highlight the pivotal role of Quantum Monte
Carlo (QMC) in generating data for constructing potential energy surfaces.\cite{ceperley_2024} 
The study suggests that stochastic noise inherent in QMC can facilitate machine
learning model training, demonstrating that models can benefit from numerous,
less precise data points. These findings are supported by an analysis of
machine learning models, where noise not only helped improve model accuracy but
also enabled error estimation in model predictions.
Similarly to QMC, our semi-stochastic formulation could take advantage of many
points computed with a low accuracy.

In this section, we discuss the application of our novel algorithm within the context of computing vibrational frequencies, specifically through the example of copper chloride (\ce{CuCl}).
A demonstrative application presented here involves the determination of the equilibrium bond length and the computation of the vibrational frequency of \ce{CuCl} using the CCSD(T)/cc-pVQZ level of theory.
The procedure involves determining the CCSD(T) potential energy curve for \ce{CuCl}, followed by its analytical representation through a Morse potential fitting:
\begin{equation}
E(r) = D_e \left( 1 - e^{-a (r - r_e)} \right)^2 + E_0
\end{equation}
where $E(r)$ represents the energy at a bond length $r$, $D_e$ the depth of the potential well, $r_e$ the equilibrium bond length, $a$ the parameter defining the potential well's width, and $E_0$ the energy at the equilibrium bond length. The vibrational frequency, $\nu$, is derived as follows:
\begin{equation}
\nu = \frac{1}{2 \pi c} \sqrt{\frac{2D_e a^2}{\mu}}
\end{equation}
with $\mu$ denoting the reduced mass of the \ce{CuCl} molecule, and $c$ the speed of light.

\begin{figure}[tb]
\includegraphics[width=\columnwidth]{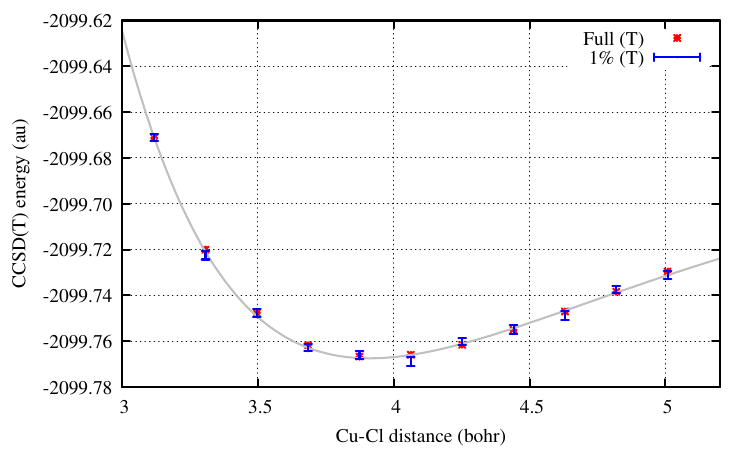}
\caption{\label{fig:cucl} CCSD(T) energies of CuCl obtained with the exact CCSD(T) algorithm (stars), the stochastic algorithm using only 1\% of the contributions (error bars), and the Morse potential fitting the points obtained with the stochastic algorithm.}
\end{figure}

The initial step involved the precise calculation of the CCSD(T) energy across various points along the potential curve.
We froze the six lowest molecular orbitals, specifically the $1s$ orbital of \ce{Cl} and the $1s$, $2s$, and $2p$ orbitals of \ce{Cu}, and correlated 34 electrons within 157 molecular orbitals.
The fitted Morse potential revealed a vibrational frequency of $\nu = \SI{414.7}{\per\centi\meter}$ and an equilibrium bond length of $r_e = \SI{3.92}{\bohr}$, aligning remarkably well with experimental values from the NIST database\cite{nist_2022} $\nu = \SI{417.6}{\per\centi\meter}$ and $r_e = \SI{3.88}{\bohr}$.

Subsequently, we applied our semi-stochastic algorithm to estimate the perturbative triples correction, utilizing merely 1\% of the total contributions.
This approach yielded a hundredfold acceleration in computational efficiency, achieving statistical uncertainty within the range of \SI{1.2} to \SI{2.0}{\milli\hartree} for each data point.
The vibrational frequency and equilibrium distance estimated using this data, $\nu = \SI{415.1}{\per\centi\meter}$ and $r_e = \SI{3.91}{\bohr}$, demonstrated comparable precision to the full computational results.
Figure \ref{fig:cucl} illustrates the potential energy surface of \ce{CuCl}, displaying both the exact CCSD(T) energies and those estimated via the semi-stochastic method.

\subsection{Performance analysis}

The primary bottleneck of our proposed algorithm lies in the generation of the sub-tensor $W^{abc}$ for each $(a,b,c)$ triplet, as discussed in Section~\ref{sec:theory}.
However, we have outlined a strategy to reframe this operation into BLAS matrix multiplications,\cite{form_w_abc} offering the potential for significantly enhanced efficiency.

We evaluated the efficiency of our implementation using the Likwid\cite{treibig_2010} performance analysis tool on two distinct x86 platforms: an AMD \textsc{Epyc} 7513 dual-socket server equipped with 64 cores at \SI{2.6}{\giga\hertz}, and an Intel Xeon Gold 6130 dual-socket server with 32 cores at \SI{2.1}{\giga\hertz}.
We linked our code with the Intel MKL library for BLAS operations.
Additionally, we executed the code on an ARM Q80 server featuring 80 cores at \SI{2.8}{\giga\hertz}, and although performance counters were unavailable, we approximated the Flop/s rate by comparing the total execution time with that measured on the AMD CPU.
On the ARM architecture, we utilized the \textsc{ArmPL} library for BLAS operations.

\begin{table*}[htb]
\begin{ruledtabular}
\begin{tabular}{lcccccc}
CPU & $N_{\text{cores}}$ & $V$ & $F$   & Memory Bandwidth & Peak DP   & Measured performance \\
               &         &     & (GHz) &      (GB/s)      & (GFlop/s) & (GFlop/s) \\
\hline
\textsc{EPYC} 7513      &      64 &  4  &  2.6  &    409.6         &     2~662 & 1~576 \\
Xeon Gold 6130 &      32 &  8  &  2.1  &    256.0         &     2~150 &   667 \\  
ARM Q80        &      80 &  2  &  2.8  &    204.8         &     1~792 &   547 \\  
\end{tabular}
\end{ruledtabular}
\caption{\label{tab:flops} Average performance of the code measured as the number of double precision (DP) floating-point operations per second (Flop/s) on different machines.}
\end{table*}

Table~\ref{tab:flops} summarizes the performance tests.
Peak performance is determined by calculating the maximum achievable Flops/s on the CPU using the formula:
\begin{equation}
P = N_{\text{cores}} \times N_{\text{FMA}} \times 2 \times V \times F
\end{equation}
where $F$ represents the processor frequency, $V$ the number of double precision elements in a vector register, $N_{\text{FMA}}$ denotes the number of vector fused multiply-accumulate (FMA) units per core (all considered CPUs possess two), and $N_{\text{cores}}$ reflects the number of cores. Notably, the Xeon and ARM CPUs both operate at approximately 30\% of peak performance, while the AMD \textsc{Epyc} CPU demonstrates twice the efficiency, achieving 60\% of the peak.

The relatively modest performance, at around 30\% efficiency, is attributed to the small dimensions of the matrices involved.
The largest matrix multiplications in the computational task entail a matrix of size ${N_\text{o}}^2 \times N_\text{v}$ and another of size $N_\text{v} \times N_\text{o}$ to yield an ${N_\text{o}}^2 \times N_\text{o}$ matrix.
These multiplications exhibit an arithmetic intensity of
\begin{equation}
I = \frac{2\, {N_\text{o}}^3\, N_\text{v}}{8\, \qty({N_\text{o}}^3 + {N_\text{o}}^2 N_\text{v} + {N_\text{o}} N_\text{v})}
\end{equation}
which can be approximated by $N_\text{o} / 4$ flops/byte as an upper bound, which is usually relatively low.
For instance, in the case of benzene with a triple-zeta basis set, the arithmetic intensity is calculated to be 3.33 flops/byte, falling short of the threshold required to attain peak performance on any of the CPUs.
By leveraging memory bandwidth and double precision throughput peak, we determined the critical arithmetic intensity necessary to achieve peak performance. On the Xeon and ARM CPUs, this critical value stands at approximately 8.4 and 8.8 flops/byte, respectively. Meanwhile, the \textsc{EPYC} CPU exhibits a value of 6.5 flops/byte, thanks to its superior memory bandwidth.

\subsection{Parallel efficiency}

\begin{figure}[tb]
\includegraphics[width=\columnwidth]{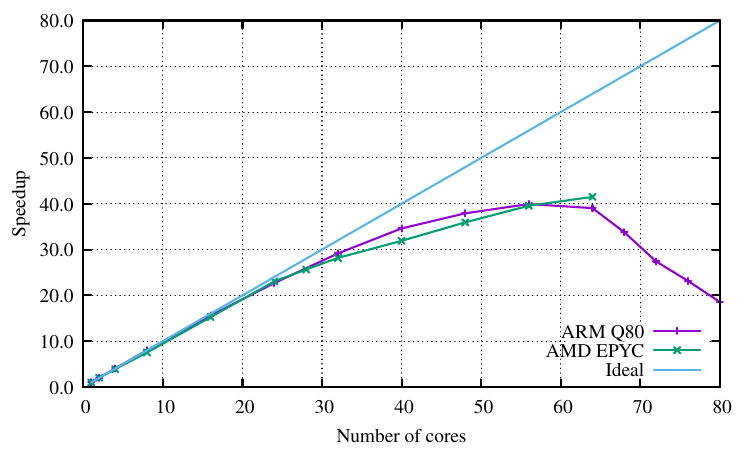}
\caption{\label{fig:speedup} Parallel speedup obtained with the ARM Q80 and AMD \textsc{Epyc} servers.}
\end{figure}

The parallel speedup performance of the ARM and AMD servers for computations involving the benzene molecule in a triple-zeta basis set is illustrated in Figure~\ref{fig:speedup}. The results delineate three distinct performance regimes:
\begin{itemize}
\item In the first regime, encompassing up to 24 cores, the performance closely approximates the ideal, with nearly linear speedup.
\item The second regime, spanning 24 to 64 cores, shows decent performance, achieving a 40-fold acceleration with 64 cores.
\item The third regime begins beyond 64 cores, where parallel efficiency rapidly deteriorates.
\end{itemize}

This performance behavior can largely be attributed to the arithmetic intensity and the bandwidth characteristics of these servers.
On the ARM server, the peak performance is attained at an arithmetic intensity of 8.75~flops/byte.
Notably, with fewer cores, the bandwidth per core increases, thereby enhancing efficiency.
For the benzene molecule in the triple-zeta basis set, the critical arithmetic intensity is 3.33~flops/byte.
This intensity corresponds to a threshold of approximately 30 cores for the ARM server and 32 cores for the AMD server.
Beyond these thresholds, particularly after 64 cores on the ARM server, the heavy demand on memory bandwidth results in a rapid decline in speedup.


\section{Conclusion}
\label{sec:conclusion}

In this work, we introduced a semi-stochastic algorithm for accelerating the computation of the perturbative triples correction in coupled cluster calculations.
This novel approach combines deterministic and stochastic methods to optimize both accuracy and computational efficiency.
The core of our algorithm is based on selectively calculating contributions labeled by triplets of virtual orbitals leveraging Monte Carlo sampling, and employing memoization to suppress redundant calculations.

Our results demonstrate that the semi-stochastic algorithm substantially reduces the computational effort compared to traditional deterministic methods, achieving near-exact accuracy with significantly reduced computational resources. Specifically, we have shown that the algorithm can achieve chemical accuracy with a small fraction of the computational effort required by fully deterministic approaches. This efficiency opens up new possibilities for studying larger systems or employing more extensive basis sets that were previously beyond reach due to computational constraints.
Additionally, the implementation of this algorithm has proven to be highly parallelizable, demonstrating excellent scalability across different platforms.

An important aspect of our investigation focused on the application of our algorithm to potential energy surface scanning.
Our method aligns well with recent findings suggesting the utility of numerous, less precise data points in constructing machine learning models.\cite{ceperley_2024}
For instance, we demonstrated that fitting a PES using data points generated with relatively large error bars using our algorithm still resulted in highly accurate values for the vibrational frequency and the equilibrium distance of copper chloride.
This capability to produce large datasets with controlled accuracy efficiently will be particularly advantageous for training machine learning models that are robust to variations in input data quality.

Therefore, our semi-stochastic algorithm not only addresses the challenge of computational expense in quantum chemistry calculations but also facilitates the generation of extensive datasets needed for machine learning applications.
This method holds significant potential for advancing computational studies in chemistry, particularly in dynamic simulations and large-scale electronic structure investigations. We advocate for continued exploration of this methodology to expand its application to other computationally demanding tasks in quantum chemistry and to explore further integration into machine learning-based chemical research.

\acknowledgements{%
  The authors kindly acknowledge fruitful discussions with Andreas
Gr\"{u}neis, Andreas Irmler and Tobias Sch\"{a}fer.
This work was supported by the European Centre of Excellence in
Exascale Computing TREX --- Targeting Real Chemical Accuracy at the
Exascale.
This project has received funding from the European Union's Horizon
2020 — Research and Innovation program --- under grant agreement
No.~952165.
Y. Damour acknowledges support and funding from the European Research Council (ERC) (Grant Agreement No.~863481).
A. Gallo acknowledges support and funding from the European Research Council (ERC) (Grant Agreement No.~101087184).
This work was performed using HPC resourced from CALMIP (Toulouse)
under allocations p18005 and p22001.%
}

\section*{Data availability statement}
The data used to produce all the plots is available at \url{https://zenodo.org/doi/10.5281/zenodo.11302501}.

\bibliography{stochastic_triples}

\end{document}